\documentstyle[prl,aps]{revtex}
\newcommand{\be}{\begin{equation}}
\newcommand{\ee}{\end{equation}}
\begin{document}
\title{Particle unstable nuclei in the Hartree-Fock theory}

\author{A. T. Kruppa }
\address{
Institute of Nuclear Research of the Hungarian Academy of Sciences,
H-4001 Debrecen, Pf. 51, Hungary}
\author{P.-H. Heenen}
\address{Service de Physique Nucl\'eaire Th\'eorique,
U.L.B.-C.P.299, B-1050 Brussels, Belgium}
\author{H. Flocard}
\address{Division de Physique Th\'eorique, Institut de Physique Nucl\'eaire,
91406 Orsay Cedex, France}
\author{R. J. Liotta}
\address{Manne Siegbahn Institute of Physics, Frescativagen 24,
S-10405 Stockholm, Sweden}

\maketitle

\begin{abstract}

Ground state energies and decay widths  of
particle unstable nuclei are calculated within
the Hartree-Fock approximation by performing
a complex scaling of the many-body Hamiltonian.
Through this transformation, the wave functions
of the resonant states become square integrable.
The method is implemented with Skyrme effective interactions.
Several Skyrme parametrizations are tested on four
unstable nuclei: $^{10}He,\ ^{12}O,\ ^{26}O$ and $^{28}O$.
\end{abstract}

{\twocolumn}
\narrowtext

The drip lines of the nuclear chart have been mapped experimentally
up to Z=19 on the proton rich side  and to Z=7 on the
neutron side\cite{conf2}.
The availability of radioactive ion beams will soon allow an
extension of our knowledge on these limits of
stability.
It will give access to the study of
nuclei beyond the drip
lines which are unstable by delayed particle emission
but have life times long enough to determine some of
their spectroscopic properties.
One-proton radioactivity has been discovered long ago\cite{Det89}.
Very recently,
two-proton emission from an unbound ground state has been
observed in the three-body decay of $^{12}O$ \cite {Kry95}.
The mass and excited states of the long searched heavy helium isotope
$^{10}He$  (which  has  the largest observed  neutron  to  proton
ratio) has  been measured \cite{Ost94,Boh95} using  a beam
where both the projectile and the target are radioactive nuclei.

The  description of  nuclei
close to the drip lines  has to deal with many specific difficulties.
In mean field methods, these difficulties are
related to the fact that the Fermi level is close to the continuum.
As a result, the last occupied orbitals have
long tails not easily described by conventional
expansions on an oscillator basis.  Pairing correlations
also create an interaction between bound and continuum
single particle levels
which cannot be treated within the BCS approximation.
These processes affect the shell structure of these isotopes
and modify the magic numbers close to the drip lines.
A proper way to solve these problems within the mean-field approximation
is the use of the Hartree-Fock-Bogoliubov
approximation with a solution of the HFB equations on a mesh\cite{DFT84}.
This method has recently been  implemented for
spherical nuclei\cite{Naz96a} and in a slightly different
way for nuclei with quadrupole triaxial deformations\cite{Ter96}.

A further complication arises for nuclei beyond
the drip lines for which the last particle orbitals are unbound
and have a width.
This is the problem that we address
in this paper.

Radioactive decay has been the subject of
numerous theoretical studies since the beginning
of nuclear physics.
The most successful approaches have been based
on the time independent formulation of nuclear collisions
and rely on an analysis of the collision matrix.
In the R matrix formulation\cite{La58},
the configuration space is divided into
an interaction region and an asymptotic
one corresponding to the daughter nuclei.
Analysis of the R matrix and the use of its relation
to the collision matrix permits to define
the energies and widths of resonances
narrow enough not to interfere in a too complicated way.
This method has been mainly used to analyze
experimental data or within phenomenological models.
Recently, a parameter free description of $\alpha$ radioactivity
based on a unification of the shell and the cluster models
has been applied to the decay of $^{212}$Po\cite{Var92}.
For light nuclei, the microscopic R matrix method
enables also to determine the decay width of unstable
states in a purely microscopic approach\cite{De96}.

Decay widths can be roughly explained
by  considering  only the penetrability through the nuclear, Coulomb  and
centrifugal  barriers, as Gamow did in the beginning  of  quantum
mechanics. This was recently done \cite{Naz96a} in a study of
the  diproton decay of $^{48}Ni$ with the  additional  assumption
that the diproton moves in the self-consistently
determined field of $^{46}Fe$.

In this letter we present a method suitable for the
calculation of the total
decay width of an unstable nucleus. It is based on
a general technique to find resonance states:
the complex scaling (CS) method.
This method  can
be combined with any model that defines
the spatial coordinates of the particles.
It has already been applied in nuclear physics
\cite{Kru,Cso,japan} to different variants
of the cluster model.
The combination of the CS and the
Hartree-Fock (HF) methods that we  use here has already
been performed
in atomic and molecular physics (for a recent reference,
see for instance Ref.\cite {Ben90}).

The starting point of the CS method is a
transformation of the Hamiltonian $\hat H$. First one defines
the unbounded non-unitary scaling operator $\hat U(\theta)$:
\be
\hat U(\theta)\Psi (\vec r)=exp(3i\theta/2) \Psi (\vec r exp(i\theta)),
\ee
where $\theta$ is real. The transformed
complex scaled Hamiltonian is of the form:
\be
\hat H_\theta=\hat U(\theta)\hat H \hat U(\theta)^{-1},
\ee
where $\theta<\pi/4$ is the scaling parameter.
If we deal with a many body system each single-particle
coordinate has to be transformed according to this prescription.

The ABC theorem\cite{Agu71} gives the properties of the spectrum
of the transformed Hamiltonian.
A bound state eigenvalue of $\hat H$ remains also an eigenvalue of
$\hat H_\theta$. A resonance pole
${\cal E}=E-i\Gamma/2$ of the Green-operator of $\hat H$
becomes an eigenvalue of $H_\theta$ provided
$\theta> {\rm arg}({\cal E})/2$. There are no other eigenvalues
of $\hat H_\theta$.
The important point is that now the wave functions of
resonant states are square
integrable. The continuous part of the spectrum of $\hat H_\theta$
differs drastically from that of $\hat H$. It is rotated down into
the complex energy plane by the angle $2\theta$.

Complex scaling can be combined with the HF theory
when resonances are to be included in the formalism. In such a
situation two cases can be
distinguished. The first one corresponds to a nucleus
with a particle stable ground state. The CS permits then
to determine the resonant states corresponding to the  self-consistent
mean field.
An RPA theory using such resonant states was recently
developed and successfully
applied \cite{TV} starting from a phenomenological  mean-field.

The second case concerns nuclei with particle unstable
ground states which therefore are resonant states.
In these cases, the CS of the many-body Hamiltonian
has to be carried out at the very first place.
This defines $\hat H_\theta$.
Choosing a Slater-determinant as trial wave function and
applying a bi-variational principle to get the ``best" single-particle
orbits, one derives new mean field equations.
The use of a bi-variational
principle \cite {Low83} 
instead of the usual Rayleigh-Ritz variational principle
is due to the non-self-adjoint nature of $\hat H_\theta$.
It was prooved that the new mean-field equation is identical with
the complex scaling of the original Fock-operator \cite{Low89}.
The total binding energy
${\cal E}_{CSHF}=E-i{{\Gamma}\over{2}}$
resulting from the application of the complex
scaled Hartree-Fock (CSHF)
procedure is complex. Its interpretation
in the many-body case is
the same as for a two-body
resonance. The half-life time of the ground state is
$\hbar\ln\, 2/\Gamma$ and the real part of ${\cal E}_{CSHF}$
is interpreted as the total binding energy which
has to be compared with the experimental mass.

We have constructed a spherical CSHF code
for Skyrme like effective interactions.
Let's consider an even-even doubly closed shell nucleus
with time-reversal invariance and assume spherical symmetry. In this case 
the HF integro-differential equation becomes an ordinary second order
differential equation
\begin{eqnarray}
{{\hbar^2}\over {2m(r)}}&&
\left(-R^{\prime\prime}(r)-{{2}\over {r}}R^\prime(r)+
{{l(l+1)}\over{r^2}}R(r)\right)\nonumber\\
&&-{{\hbar^2}\over {2m^\prime(r)}}R^\prime(r)+V(r)R(r)=\epsilon R(r),
\end{eqnarray}
where $R(r)$ is the radial part of the single-particle orbit.
The effective mass $m(r)$ and the
mean field potential $V(r)$ are complicated functionals
of the filled single particle orbits. The actual form of these
functionals are given in \cite{DFT84}.
The left hand side of Eq. 3 is actually the action
of the Fock-operator $\hat h$ onto the function $R(r)$.
In order to find the resonances one has to derive 
the complex scaled Fock-operator 
$\hat h_\theta=\hat U(\theta) \hat h \hat U(\theta)^{-1}$, 
which is a rather straightforward task.
The action of $\hat h_\theta$
on an arbitrary function $f(r)$ is
\begin{eqnarray}
\hat h_\theta f(r)={{\hbar^2}\over {2m(r\eta)}}&&
\eta^{-2}\left(-f^{\prime\prime}(r)-{{2}\over {r}}f^\prime(r)+
{{l(l+1)}\over{r^2}}f(r)\right)\nonumber\\
&&-\eta^{-1}{{\hbar^2}\over {2m^\prime(r\eta)}}f^\prime(r)+V(r\eta)f(r),
\end{eqnarray}
where $\eta=\exp (i\theta)$. The eigenvalue problem of $\hat h_\theta$
is solved by direct numerical integration and the usual
iteration technique is used except that the initial orbits were
calculated using a complex scaled radial Schr\"odinger equation.
Further details of the present calculation and a discussion on the
choice of the 
basis set used in the CSHF approach
will be published elsewhere.

We will apply the CSHF method
to calculate the widths of the ground state
of the nuclei $^{10}He,\ ^{12}O,\ ^{26}O$ and $^{28}O$.
The parameters of  effective nucleon-nucleon interactions
are fitted to reproduce properties of selected stable nuclei
as well as nuclear and neutron matters.
These interactions are often more reliable than
phenomenological approaches in the determination
of nuclear properties far from stability.
However, the isospin dependence of the interaction
is probably rather poorly known.
In this work, we not only determine the energies
of nuclei close to the drip lines but we also
calculate the width of resonant states,
a property which has not at all been included
in the adjustment of the forces.
We will  therefore test farther a large number of
interactions which have been rather successful
in the description of a large variety of nuclear properties.
The parametrization SIII has been the more widely used,
in particular to describe deformation properties of long series
of isotopes.  The interaction SLy4 \cite{Cha96} 
has been recently adjusted with a special
care on the properties of nuclear and neutron matters and is expected
to be particularly reliable far from stability. The force 
Skm1 \cite{Gom95} is a slightly modified version of Skm*. 
It should have the
same properties close to stability line and be well suited to
describe extremely deformed nuclear states.
However, the very bad isospin dependence of Skm*
has been corrected by a reduction of its symmetry energy
from 30.0MeV to 25.8MeV. Although this value seems to be
too small, Skm1 leads to very good prediction on masses
of nuclei far from stability in the HF approach. Finally, we have also
used the parametrization SkP \cite{DFT84}
which has been constructed to
be used at the same time in the mean field and in the pairing channels.
This interaction is the only one to have an effective mass
equal to 1, and therefore produces a larger density of levels
in the vicinity of the Fermi energy. This difference could
be important for nuclei unstable by particle emission.

In table I the  absolute value of the binding energies of  even-even
$He$ and $O$ isotopes and of  $^{10}C$ are given.
In the case of a partially filled shell,
the uniform filling approximation has been used.
For most calculations, the Fermi level of the nuclei were found to be
bound and the standard HF procedure has been applied.
In the case  of the Skm1 parametrization,
the Fermi energy is in the continuum for
 $^{10}He,\ ^{12}O,\ ^{26}O$ and $^{28}O$
and for SIII in the case of $^{10}He$.
The CSHF method was used in these circumstances and
the absolute value of the real part of the total binding energy
is displayed in Table I. In order to test the accuracy of the
the CSHF method, we have carried out CSHF calculations
in a few bound isotopes. Because of the finite accuracy of the
numerical integration of the differential equation (4)
the total energy has a
small imaginary part. In each cases,
its absolute value was less than $10^{-4} MeV$.
This gives an estimation of the overall error
and indicates
that we could not attempt to describe a system with a decay
width smaller than $10^{-4} MeV$.

On the average, the parametrization Skm1  leads to
the best agreement with the experimental energies,
with the noticeable exception of $^{10}He$.
 The fact that Skm1 leads to a
very fast decrease of the binding energies for the most
neutron rich isotopes may probably be related to its
behaviour in pure neutron matter. Compared to realistic
calculations, this parametrization leads to an overestimation
of the energy of neutron matter at densities lower than
0.75 times the saturation density and to a collapse
for larger densities\cite{Mey96}. The underestimation
of the energy of $^{10}He$ is thus probably related
to the presence of regions of low densities due to the
large neutron excess. Note that the interaction Skm*
from which Skm1 is derived describes much more satisfactorily
neutron matter.
Differences between Skm1, SIII and Sly4 are
large only for the most neutron rich isotopes
while systematically larger differences 
are obtained with SkP. Note
that when pairing correlations are included\cite{Dob96}, SIII, Sly4
and SkP predict $^{26}O$ to be bound by at least 2.0 MeV
and $^{28}O$ to be either slightly unbound (SIII) or
slightly bound.

In table II we compare the CSHF  ground state decay widths
to the  experimental data.
The four forces predict  $^{10}He$ to be unbound
with respect to two proton emission. However
with  SkP and SLy4, all single particle orbits are bound.
The Fermi level lies in the continuum for SIII and Skm1 which
makes possible the calculation of the decay width of
$^{10}He$ by the CSHF method.
The SIII width is in good agreement with the experimental data.
The values obtained with Skm1 overestimate the experimental
data by an order of magnitude for $^{10}He$ while it largely
underestimates it for $^{12}O$. The reasons for these discrepancies
are probably different. $^{10}He$ is largely under-bound by Skm1
and more correctly described by SIII. The too large width obtained
sith Skm1 is probably related to this under-binding.
 $^{12}O$ is not a closed shell nucleus and is described by
the filling approximation. A more realistic
description of this nucleus would affect the decay width.
Let us also recall that correlations beyond the mean field approximation
are important in  light nuclei.

The forces Skm1 and SIII predict that both $^{26}O$ and $^{28}O$ are
unbound with respect to neutron emission.
The interaction Sly4 also gives unbound $^{26}O$.
However only the parametrization Skm1 places the Fermi-level in the
continuum  so that the decay width can be calculated by the CSHF method.
Looking to the results on $^{12}O$ and $^{10}He$, one can expect that
the  widths given in Table II are more reliable for $^{28}O$,
which is a closed shell nucleus than for $^{26}O$ for which the
filling approximation has been used.

In conclusion, we have shown in this paper
that nuclei with a particle unstable ground state
can be described by the complex scaled Hartree-Fock method.
This method permits to determine
the energies and the widths for particle
emission, while the conventional HF method does not even
give access to energies. Our results on $^{10}He$ have
revealed a weakness of the interaction Skm1
which was not apparent on stable
nuclei. While the feasibility of the method has
clearly been demonstrated, the values of widths that we
have obtained are not in agreement with the experimental
data. A part of the discrepancy is probably due to the
fact that we use effective nuclear interactions in a domain
for which they have not been constructed. Before to address
this problem, we plan to generalize our method
by the inclusion of pairing correlations.

We thank J. Dobaczewski and J. Meyer for interesting
discussions.
ATK is grateful for support from the Hungarian OTKA Grant No. T17298.
This work has been partly supported by the Belgian
SSTC under contract ARC 93/98-166.

\widetext
\begin{table}
\caption{The negative of the binding energies of light nuclei in MeV with
different Skyrme forces using the HF or CSHF method.}
\begin{tabular} {lccccccccccc}
&$^6He$&$^8He$&$^{10}He$&$^{10}C$&$^{12}O$&$^{14}O$
&$^{16}O$&$^{22}O$&$^{24}O$&$^{26}O$&$^{28}O$ \\
\hline
Exp.\tablenote{See the Ref.
\cite{Aud93}.}&29.3&31.4&30.3&60.3&58.5&98.7&127.6&162.0&168.5&168.4&\\
Skm1&29.1&31.0&25.0&60.9&58.4&101.4&127.7&164.0&171.9&167.9&164.6\\
SIII&29.3&33.6&29.5&59.4&59.1&100.9&128.2&165.4&171.8&170.9&171.4\\
Sly4&29.5&33.9&31.4&59.5&60.9&101.5&128.5&164.4&172.4&172.3&173.8\\
SkP&30.5&34.7&32.7&60.7&61.9&101.2&127.6&164.4&173.5&174.0&175.1\\
\end{tabular}
\label{binding}

\end{table}

\narrowtext
\begin{table}
\caption{The decay width of the ground state of $^{10}He,\ ^{12}O,\ ^{26}O$
and $^{28}O  $
in $MeV$ using the CSHF model.
With the other forces of Table I. decay
width could not be assigned to these nuclei.}
\begin {tabular} {lllll}
&$^{10}He$&$^{12}O$&$^{26}O$&$^{28}O$\\
\hline
Exp.&0.3\tablenote{See the Ref. \cite{Boh95}.}&0.578\tablenote{See the Ref.
\cite{Kry95}.}&& \\
Skm1&3.8&$<10^{-3}$&0.4&0.5 \\
SIII&0.4&&&\\
\end{tabular}
\label{gamma}
\end{table}

\end{document}